# Assessment of the probability of microbial contamination for sample return from Martian moons I: Departure of microbes from Martian surface

Kazuhisa Fujita[a], Kosuke Kurosawa[b], Hidenori Genda[c], Ryuki Hyodo[c], Shingo Matsuyama[d], Akihiko Yamagishi[e], Takashi Mikouchi[f], and Takafumi Niihara[g]

[a] Institute of Space and Astronautical Science, Japan Aerospace Exploration Agency, 3-1-1, Yoshinodai, Chuo-ku, Sagamihara, Kanagawa 252-5210, Japan.
[b] Planetary Exploration Research Center, Chiba Institute of Technology, 2-17-1, Narashino, Tsudanuma, Chiba 275-0016, Japan.
[c] Earth-Life Science Institute, Tokyo Institute of Technology, 2-12-1 Ookayama, Meguro-ku, Tokyo 152-8550, Japan.
[d] Aeronautical Technology Directorate, Japan Aerospace Exploration Agency, 7-44-1, Jindaijihigasi-machi, Chofu, Tokyo 231-0858, Japan.
[e] Department of Applied Life Sciences, School of Life Sciences, Tokyo University of Pharmacy and Life Sciences, 1432-1, Horinouchi, Hachioji, Tokyo 192-0392, Japan
[f] The University Museum, The University of Tokyo, 7-3-1, Hongo, Bunkyo-ku, Tokyo 113-0033, Japan
[g] Department of Systems Innovation, School of Engineering, The University of Tokyo, 7-3-1 Hongo, Bunkyo-ku, Tokyo 113-8656, Japan

Corresponding author :
Kazuhisa Fujita, Professor
Institute of Space and Astronautical Science, Japan Aerospace Exploration Agency, 3-1-1, Yoshinodai, Chuo-ku, Sagamihara, Kanagawa 252-5210, Japan.
Email: fujita.kazuhisa@jaxa.jp
Tel: +81-50-3362-4378; Fax: +81-422-40-3245; ORCID ID: 0000-0001-9589-4913

Running title :
Departure of microbes from Martian surface

**Abstract**

Potential microbial contamination of Martian moons, Phobos and Deimos, which can be brought about by transportation of Mars ejecta produced by meteoroid impacts on the Martian surface, has been comprehensively assessed in a statistical approach, based on the most probable history of recent major gigantic meteoroid collisions on the Martian surface. This article is the first part of our study to assess potential microbial density in Mars ejecta departing from the Martian atmosphere, as a source of the second part where statistical analysis of microbial contamination probability is conducted. Potential microbial density on the Martian surface as the source of microorganisms was estimated by analogy to the terrestrial areas having the similar arid and cold environments, from which a probabilistic function was deduced as the asymptotic limit. Microbial survival rate during hypervelocity meteoroid collisions was estimated by numerical analysis of impact phenomena with and without taking internal friction and plastic deformation of the colliding meteoroid and the target ground into consideration. Trajectory calculations of departing ejecta through the Martian atmosphere were conducted with taking account of aerodynamic deceleration and heating by the aid of computational fluid dynamic analysis. It is found that Mars ejecta smaller than 0.03 m in diameter hardly reach the Phobos orbit due to aerodynamic deceleration, or mostly sterilized due to significant aerodynamic heating even though they can reach the Phobos orbit and beyond. Finally, the baseline dataset of microbial density in Mars ejecta departing for Martian moons has been presented for the second part of our study.

**Abbreviations**

CFD    Computer Fluid Dynamics
CFU    Colony Forming Unit
SPH    Smoothed Particle Hydrodynamics

## 1. Introduction

The planet Mars and its moons, Phobos and Deimos, have attracted continual scientific interests from viewpoints of the process of chemical evolution and the origins of life. Motivated by such scientific interests, Japan Aerospace Exploration Agency (JAXA) is currently entertaining a sample return mission from Martian moons for launch in 2024 (e.g., Fujimoto, 2017). In the exploration of solar system bodies, it is of primary importance to comply with the COSPAR Planetary Protection Policy (PPP). Especially in sample return missions, it is strongly requested to protect Earth and its biosphere from potential harmful extraterrestrial sources of contamination. Although it is still a subject of controversy whether Mars was once warm and an ocean existed in the ancient time (e.g., Citron et al., 2018), it is not possible at this moment to completely deny the possibility that life was created on Mars and survived even until now. Therefore, there is a potential risk of microbial contamination of the Martian moons due to mass transportation from Mars. It is highly recommended by the COSPAR PPP that, in sample return missions from the solar system bodies having a potential risk of extraterrestrial microbial contamination, the probability that a single unsterilized particle of 10-nm diameter or grater is in a sample returned from extraterrestrial bodies shall be less than $1 \times 10^{-6}$. This is so-called the REQ-10 requirement.

In the preceding study of SterLim (ESA contract no. 4000112742/14/NL/HB) (Patel et al., 2018; Summers, 2017), it was pointed out that the REQ-10 requirement can be violated with a high probability in sample return missions from the Martian moons. However, in their original article, Mars ejecta fragments were assumed to be homogeneously distributed on the surface of the Martian moons by averaging the incoming flux for time. For this reason, in this study, we consider more realistic processes of mass transportation from Mars to the Martian moons and distribution of Mars ejecta fragments on the surface of the Martian moons, based on the most probable history of recent major gigantic meteoroid collisions, which is evident from observation of the present Martian surface. The purpose of this study is to clarify the potential physical processes which can bring about microbial contamination on the surface of Martian moons, to obtain a quantitative estimate of the density of microorganisms still surviving in the regolith of the Martian moons regardless of several sterilization processes, and to assess microbial contamination probability of samples collected on the surface of the Martian moons for future sample return missions from the Martian moons. This article is the first part of our study to assess potential microbial density in Mars ejecta departing from the Martian surface, as a source of the second article, hereafter referred to as Paper 2 (Kurosawa et al., submitted), where statistical analysis of microbial contamination probability is conducted.

## 2. Overview of Microbial Contamination of Martian Moons
### 2.1. Potential Scenario of Microbial Contamination

A potential scenario of microbial contamination of the Martian moons is illustrated in Fig. . Following the SterLim study, we consider that microbes, which potentially exist on the Martian surface (denoted as 1 in Fig.), are transported to the Martian moons by Mars ejecta produced by meteoroid impacts on the Martian surface (2). Some portions of microbes are annihilated through hypervelocity impacts during Mars ejecta formation (3) and aerodynamic heating acting on Mars ejecta during hypersonic flight in the Martian atmosphere (4). Mars ejecta that are potentially contaminated with Martian microbes are transported to orbits of the Martian moons, a small portion of which may impact against the Martian moons, destroyed into fragments with considerable shock-heating. Because of such shock-heating, a considerable portion of microbes contained in the Mars ejecta are annihilated (5). A certain portion of the Mars ejecta fragments are considered to remain in the craters on the surface of the Martian moons, and to be diluted with indigenous fragments, forming a collapsed lens of considerable thickness in the craters (see Paper 2 for more details). On the other hand, the rest of the Mars ejecta fragments and some portion of the indigenous fragments, both of which are small in diameter, are ejected into orbits around Mars, and finally re-impact on the global surface of the Martian moons, forming a contaminated layer of submillimeter thickness (6). These processes are illustrated in more detail in Figs. 1a and 1b of Paper 2.

Microorganisms contained in the craters and in the thin layers on the surface of the Martian moons are expected to gradually decrease with time due to sterilization by solar and galactic cosmic radiation (7). The time scale of radiation sterilization is Kyr in this region (see Paper 2 for more details). However, it should be noted that, in addition to Mars ejecta, natural meteoroids continuously impact on the surface of the Martian moons as well, creating craters, scattering fragments, and eventually forming a thick layer of regolith around the craters on the surface of the Martian moons (8). Because of this continuous phenomenon, a certain portion of the contaminated surface can be immediately covered with a thick layer of regolith before sufficient sterilization has been completed by radiation, leaving shielded areas where the microbial density decreases more slowly than in the exposed areas. The time scale of radiation sterilization is Myr in this region. These processes are illustrated in more detail in Figs. 1c and 1d of Paper 2. Above all, the surface of the Martian moons can be roughly divided into three categories (as illustrated in Fig. 1d of Paper 2):

1) Craters produced by Mars ejecta impacts, where microbial density may remain relatively high to a considerable depth for long, since radiation sterilization progresses slowly in deep regions.
2) Areas covered with thick regolith layers produced by natural meteoroid impacts, where microbial density decreases at a moderate speed by radiation sterilization.
3) Exposed common areas, where microbial density quickly decreases by radiation sterilization.

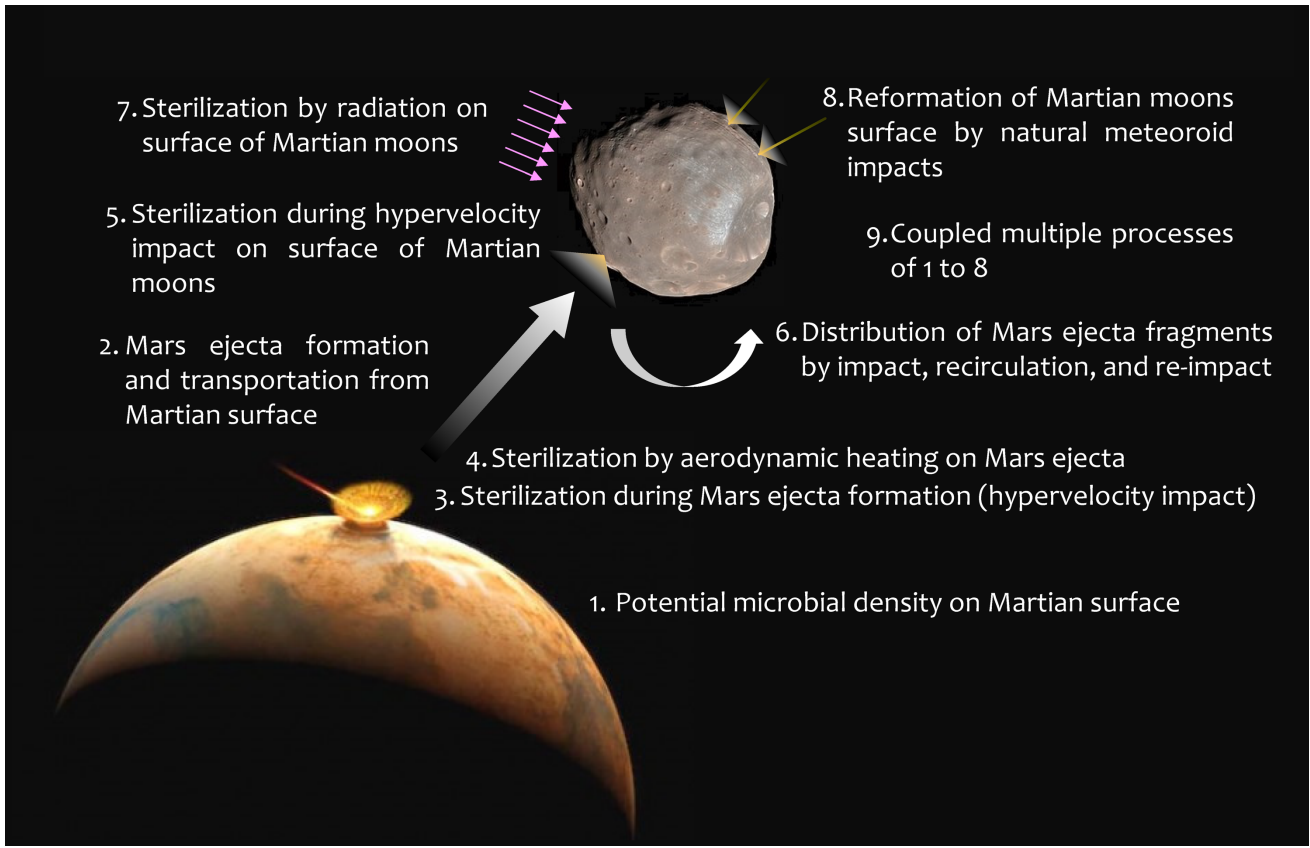

Fig. 1  Potential scenario of microbial contamination of Martian moons.

Since our study covers a lot of physical processes described above, it is divided into two articles. This article, hereafter denoted as Paper 1, deals with the processes from 1 to 4 that occur on the Martian surface, while the processes from 5 to 9 that occur on the surface of the Martian moons are described in Paper 2. As a whole, all the processes are quantitatively examined to obtain an estimate of the potential microbial density on the surface of the Martian moons, from which microbial contamination probability in samples collected on the surface of the Martian moons is assessed.

Table 1 Comparison to SterLim Study

| | Elementary processes | SterLim | Present study |
|---|---|---|---|
| 1 | Potential microbial density on Martian surface | Assuming same microbial density as Atacama Desert | Similar to SterLim, but a slight acceptable correction introduced |
| 2 | Mars ejecta formation and transportation | Based on study of Melosh | SPH computations newly conducted for statistical analysis |
| 3 | Sterilization during Mars ejecta formation | No model introduced (microbe survival rate = 1) | Sterilization during meteoroid impact additionally introduced |
| 4 | Sterilization by aerodynamic heating on Mars ejecta | N/A | Thermal analysis of Mars ejecta conducted along trajectories |
| 5 | Sterilization during hyper-velocity impact on surface of Martian moons | Microbe survival rate ~ 0.1 for velocity <2 km/s, based on impact experiments | Impact sterilization model renewed. SPH and trajectory analysis suggesting more frequent sterilization going on |
| 6 | Distribution of Mars ejecta fragments by impact, recirculation, and re-impact | Homogeneous deposition by averaging the incoming flux | Crater formation by Mars ejecta with retention & scattering of Mars ejecta fragments taken into account |
| 7 | Sterilization by radiation on Phobos surface | Sterilization model based on experimental fact | Same as SterLim, but integration in depth direction newly conducted |
| 8 | Surface reformation by natural meteoroid impacts | N/A | Continuous natural meteoroid impacts on Martian moons taken into account |
| 9 | Coupled multiple processes of 1 to 8 | N/A | Statistical analysis of microbial contamination conducted |

## 2.2. Comparison to SterLim Study

In Table 1, comparison is made between the physical processes taken into consideration and the numerical procedures used in the original study of SterLim and this study. Potential microbial density on the Martian surface, which is assumed as that of the Atacama Desert in the SterLim study, is revised in this study by analogy from the terrestrial areas having the similar arid and cold environments as the Martian surface. Mars ejecta formation and transportation are based on the results of Melosh and Chappaz in the SterLim study (Melosh, 2011; Chappaz et al., 2013), while the 3-dimensional SPH computation for oblique impacts (Genda et al., in preparation) and associated trajectory analysis (Hyodo et al., in preparation) are newly conducted in this study. Effects of sterilization during meteoroid impacts on the Martian surface and sterilization by aerodynamic heating acting on Mars ejecta are newly taken into account in this study. Microbe survival rate during hypervelocity impacts of Mars ejecta against the Martian moons is assumed uniquely as 0.1 in the SterLim study, based on the impact experiments, since impact velocities are estimated below 2 km/s. In contrast, the hypervelocity impact sterilization model is renewed and extended to a higher velocity range in this study by using the experimental data of the SterLim study, since the results of the SPH computation and the trajectory analysis suggest that impact velocities are in general higher than 2 km/s. The radiation sterilization model used in this study is taken from the SterLim study. The greatest difference in this study is that, in the SterLim study, Mars ejecta is assumed to be continuously and globally deposited on the surface of the Martian moons, whereas three-dimensional distribution of Mars ejecta fragments is considered, according to the most probable scenario of past meteoroid impact events on the Martian surface in the history of Mars.

## 3. Potential Microbial Density on Martian Surface

Estimation of microbial density on the Martian surface is of primary importance since the final result of microbial contamination probability linearly changes with the estimated microbial density. Since it is reasonable to get insight from microbial density in the terrestrial areas having the similar arid and cold environments as the Martian surface, a thorough survey of the literatures on microbial density in the analogical environments has been conducted below.

Navarro-González et al. (2003) reported microbial densities ranging from $10^5$ to $10^{10}$ CFU/kg in the Atacama Desert, as shown in Fig. 2. However, it should be noted that microbial densities measured in the Yungay area, which is the most arid zone in the Atacama Desert, are not greater than $10^7$ CFU/kg. Maier et al. (2004) reported microbial densities ranging from $10^6$ to $10^8$ CFU/kg in the Yungay area as well. Their results show good agreement with those of Navarro-González et al. (2003). Since the Martian surface is in general extremely arid as the Yungay area, it might be more reasonable to take the values measured in this area. Glavin et al. (2004) estimated bacterial cell abundances near the Yungay area, finding that bacterial counts for the surface and subsurface samples of $0.7 \times 10^9$ cells/kg and $9.6 \times 10^9$ cells/kg, respectively, by using the DAPI (4,6-diamidino-2-phenylindole) staining method. More recently, Drees et al. (2006) measured culturable biomass count of $5.40 \times 10^6$ CFU/kg in the Yungay area, and $9.11 \times 10^7$ to $1.36 \times 10^8$ CFU/kg in the surrounding area. Phospholipid fatty acids (PLFA) analysis of Lester et al. (2007) indicated $8.5 \times 10^9$ to $6.0 \times 10^{10}$ cells/kg in the Yungay area, while culturing of soil extracts on R2A and TSA media yielded $6.3 \times 10^5$ to $5.2 \times 10^6$ CFU/kg at the same time. Connon et al. (2007) reported that PLFA concentrations ranged from $2 \times 10^8$ to $7 \times 10^9$ cells/kg while R2A culturing yielded $4.7 \times 10^4$ CFU/kg in the Yungay area.

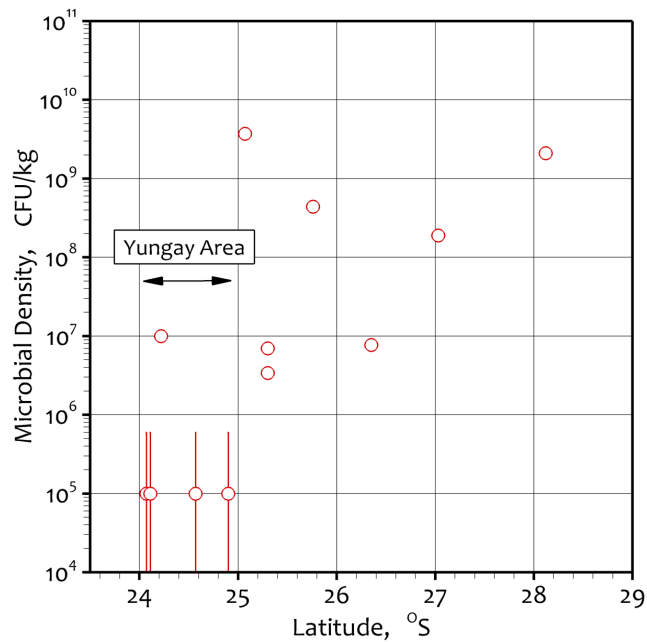

Fig. 2    Microbial density measured in the Atacama Desert (Navarro-González et al., 2003).

In addition to the Atacama Desert, microbial populations in Antarctic permafrost are considered to give closer information, since the Antarctic permafrost is more likely to bear analogy with the Martian environments due to low temperature. Aislabiea et al. (2006) reported the diversity of soil bacterial communities from four locations in Victoria Land of Antarctica, Marble Point, Bull Pass, Lake Vanda, and Mt. Fleming, whose mean annual temperature is –18, –20, –20, and –24 °C, respectively. The total direct count by molecular analysis ranged from $10^9$ to $1.3 \times 10^{11}$ cells/kg, while number of culturable heterotrophs was found to range from $10^4$ to $3.7 \times 10^8$ CFU/kg, depending on the environments. It should be noted that cell count was the lowest or below the detection limit of $10^4$ CFU/kg in the coldest areas, which are Bull Pass and Mt. Fleming in their study. Gilichinsky et al. (2007) measured the number of bacterial cells in the frozen ground and the buried soils of the McMurdo Dry Valleys in Antarctica, whose mean annual temperature is –20 °C in general, finding that the total cell number counted by epifluorescence microscopy was $10^6$–$10^9$ cells/kg. Very recently, Goordial et al. (2016) reported very low microbial biomass of $1.4$–$5.7 \times 10^6$ cells/kg in both the dry and ice-cemented permafrost of the University Valley, which is one of the coldest and driest regions in the McMurdo Dry Valleys (MDVs) of Antarctica, by using Dichlorotriazinyl Aminofluorescein (DTAF) stain. It is noted that the mean annual temperature of the University Valley is –23 °C, and no seasonal thaw of permafrost is expected. On the other hand, 2 orders of magnitude higher cell counts ($1.2$–$4.5 \times 10^8$ cells/kg) were detected in the active layer and permafrost soils from the Antarctic Peninsula, which has more moderate environments than the University Valley.

It should be noted that the microbial biomass count in cells/kg is much greater than the count in CFU/kg by the factor of 10 to $10^3$ in general, since it may include inactivated microbe cells as well. From a viewpoint of conservative assessment, it is appropriate to use the count in cells/kg rather than CFU/kg. For this reason, the microbial density is measured in cells/kg hereafter in this study. With a conservative assumption of 1 CFU/kg = $10^3$ cells/kg, the microbial density taken from the literatures is summarized in Fig. . It is clearly understood that the microbial density tends to decrease as the environments becomes more arid and colder. Since the mean annual temperature of Mars is considered to be roughly –60 °C (Williams, 2018), and since most of the Martian permafrost is not expected to thaw, it must be reasonable to consider that the microbial density on the Martian subsurface ranges from $10^6$ to $10^9$ cells/kg as a conservative estimation of the asymptotic limit.

Based on the above estimation, a probabilistic model of the Martian subsurface microbial density is formulated below. It is assumed that a normal distribution function of $\log_{10} n$ applies where $n$ is the microbial density. The center value and the standard deviation ($\sigma$) are assumed to be –7.5 and 0.5, respectively. Thus, the probabilistic function of the microbial density is given by

$$P_{d,m} = \sqrt{\frac{2}{\pi}} \exp[-2(x - 7.5)^2] \qquad (1)$$

$$x = \log_{10} n \qquad (2)$$

The probability density distribution of Eqs. (1) and (2) is plotted in Fig. .

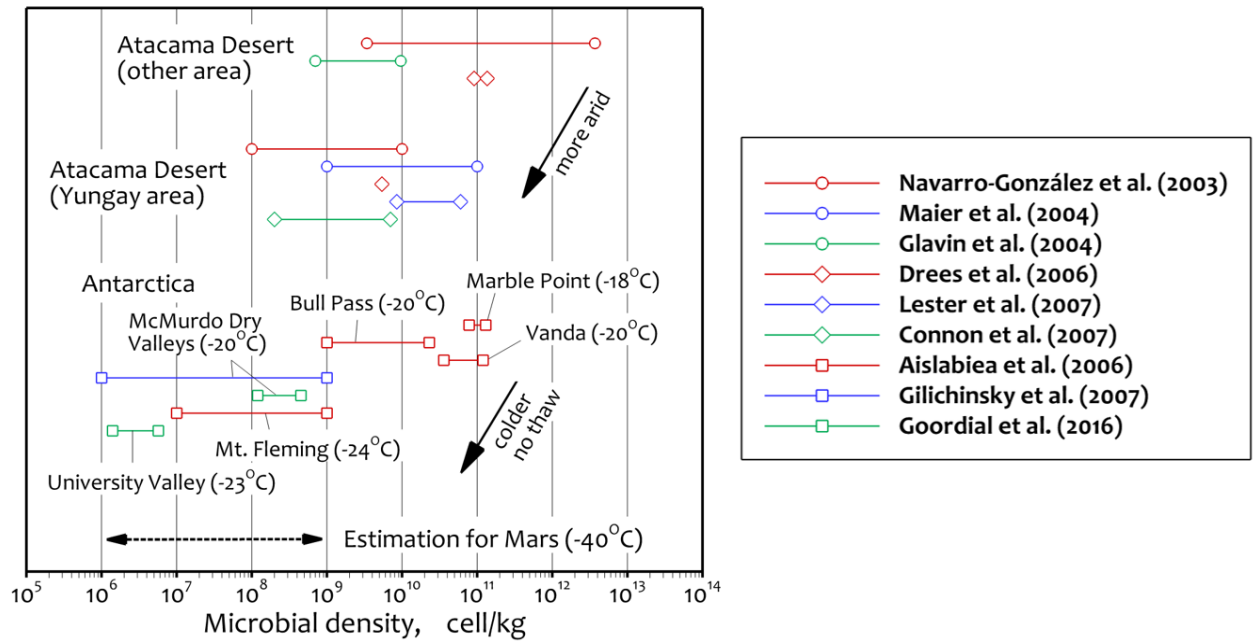

Fig. 3   Comparative summary of microbial density measured in the arid and cold environments on Earth.

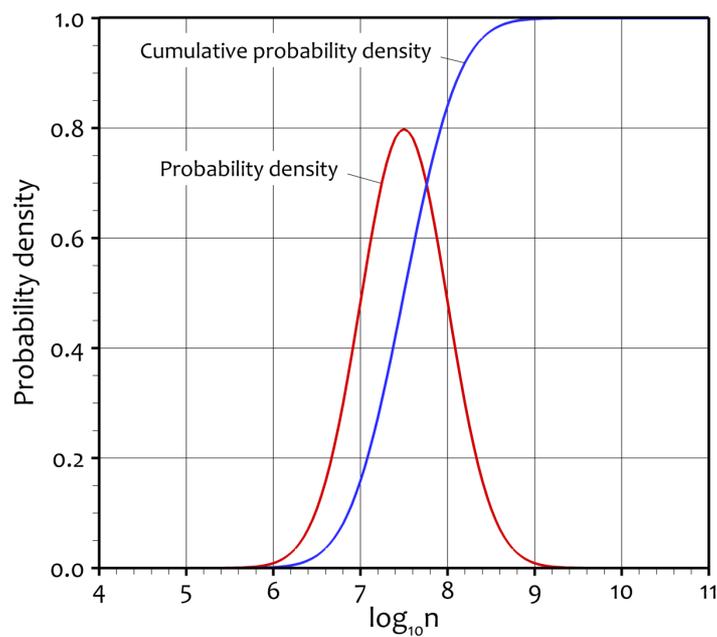

Fig. 4   Probability density function for Martian subsurface microbial density given by Eqs. (1) and (2).

Recently, Sholes et al. (2019) theoretically estimated an ideal upper limit of the total number of subsurface microbes by assuming CO– and $H_2$–consumed metabolism. Photochemical reactions in the Mars atmosphere produces out-of-equilibrium species, that is, CO and $H_2$, from $CO_2$ and $H_2O$. The free energy derived from oxidation and/or methanation of CO is expected to be the major energy source for the microbes lived near the surface. By considering several processes, including photochemical reactions, diffusion, atmospheric escape to the space, and CO and $H_2$ consumption by the metabolic activities of potential microbes, they found that the atmospheric composition near the Martian surface deviates from the measured values by the Curiosity rover when the total number of metabolizing microbes exceeds ~ $3 \times 10^{27}$ cells. They also conducted a one-dimensional diffusion calculation into the subsurface, showing that the photochemically-generated free energy could support the microbes lived down to several km form the surface. This estimate provides an upper limit of the total number of potential microbes under the subsurface condition on the current Mars because of two reasons as follows. First, they assumed all the CO and $H_2$ produced via the photochemical reactions are consumed by the microbes. Second, they employed the smallest value of the minimal basal power requirement (BPR) of organisms, which is the required energy to sustain the activities of the individual cells, reported by the previous studies. We could obtain an independent estimate of the potential microbe density from the allowable number of microbes. If we assumed the microbes of $3 \times 10^{27}$ cells homogeneously existed in the Martian crust down to 200 m, the microbe density is estimated to be ~ $4 \times 10^7$ cells/kg. This result shows satisfactory agreement with the estimated microbial density given by Eqs. (1) and (2). The mixing depth of the microbes used in the estimate corresponds to the expected excavation depth of the Zunil-forming impact event on Mars, which mainly leads to microbial contamination on each of the Martian moons (see Paper 2). Based on the observed diameter of Zunil crater and the crater scaling laws (Schmidt and Housen, 1987), we estimated the diameter of the Zunil-forming impactor to be ~2 km (Hyodo et al., in preparation). The original depth of high-speed ejecta having sufficiently high velocity to reach the orbits of the Martian moons has been estimated to be 10% of the projectile diameter (Artemieva and Ivanov, 2004; Genda et al., in preparation). This depth assumption yields a conservative estimate for the number density of the potential microbes because the microbes could live at the subsurface down to several km as mentioned above.

There is no additional scientific evidence obtained so far to determine microbial density on Martian surface. This means that the baseline value for microbial density on the Martian surface is of political nature rather than of scientific at this moment. We may have a lot of knowledge gaps on this problem. From the above discussion, it would be reasonable to assume that the baseline value of microbial density on the Martian surface is between $10^7$ to $10^8$ cells/kg ($10^{7.5 \pm \sigma}$) or less. These values are in agreement with the detection limit of organics by the Viking gas chromatography and mass spectrometry (GC/MS) experiments, which suggested that $10^9$ to $10^{10}$ cells/kg would have been undetected by these instruments (Klein, 1978; Glavin et al., 2001; Mukhopadhyay, 2007). In the following part of this article and Paper 2, we use the maximum value of $10^8$ cells/kg as the worst case, and Eqs. (1) and (2) for the comprehensive statistical analysis.

## 4. Sterilization during Mars Ejecta Formation

Mars ejecta transported to the Martian moons are produced by hypervelocity impacts of meteoroids against the Martian surface. In such the impact events that can transport the ejected rocklets to the Martian moons, the incident meteoroids must have considerably high impact velocities, since the ejected rocklets should have velocities higher than 3.8 km/s at departure from the Martian surface to reach the Phobos orbit. Because of such hypervelocity impacts, when the surface on which a meteoroid impacts is sufficiently rigid, considerable shock-heating must occur in fragmentation of the impactor and the target surface, as reported by Kurosawa and Genda (2018), resulting in high sterilization of Mars ejecta. Such the heat source due to plastic deformation of the pressure-strengthened rocks has been overlooked for a long time (Melosh and Ivanov, 2018). In order to confirm this, shock-heating generated among ejecta is numerically simulated by using the iSALE shock physics code (Amsden et al., 1980; Ivanov et al., 1997; Wünnemann et al., 2006) with taking internal friction and plastic deformation into consideration. Strength parameters are taken from the work for basalt (Ivanov et al., 2010). The numerical suite constructed by Kurosawa et al. (2018) was employed in this simulation, but we newly introduced a constitutive model for basaltic materials. The general setup parameters for the simulation and the input parameters pertaining to the material model of basalt are summarized in Tables 2 and 3, respectively. The results are shown in Fig. 5b and compared to the results without internal friction and plastic deformation shown in Fig. 5a.

Table 2 General setup parameters for the 2-D iSALE calculation. The detailed descriptions for values listed here can be found in Collins et al. (2016).

| Computational geometry | Cylindrical coordinates |
|---|---|
| Number of computational cells in $R$ direction | 2000 |
| Number of computational cells in $Z$ direction | 3000 |
| Number of cells for the extension in $R$ direction | 200 |
| Number of cells for the extension in $Z$ direction (bottom only) | 200 |
| Extension factor | 1.02 |
| Grid spacing (m/grid) | 2.5 |
| Cells per projectile radius | 1000 |
| Artificial viscosity $a_1$ | 0.24 |
| Artificial viscosity $a_2$ | 1.2 |
| Impact velocity (km/s) | 3.5 |
| High-speed cutoff (km/s) | 35 (10-fold impact velocity) |
| Low-density cutoff (kg/m$^3$) | 1 |
| Initial surface temperature (K) | 210 |
| Initial thermal structure in the target | Constant |

In Fig. 5, symbols represent the ejected particles produced in a vertical impact with an impact velocity at 3.5 km/s. The impact velocity corresponds to the normal component of the impact velocity for at 45-degrees-obluque impacts at the escape velocity of Mars. Thus, the figure shows a minimum estimate of the degree of shock heating. Color of symbols represent the ejection velocity, and the data points surrounded by black circles represents the particles having velocities higher than 3.8 km/s, which is the minimum velocity to reach the Phobos orbit. It is seen that a considerable portion of ejecta do not undergo shock-heating, remaining their temperature below 300°C (denoted by the lower dashed line in Fig. 5) when internal friction and plastic deformation are not present (Fig. 5a). In contrast to this, the ejecta having velocities greater than 3.8 km/s are heated above 1,000 K when the internal friction and plastic deformation is taken into account (Fig. 5b). This suggests that the Mars ejecta reaching the Phobos orbit and beyond cannot avoid shock-heating when the target ground is sufficiently rigid. According to the heat test results in the SterLim study, the microbial survival rate ($N/N_0$ where $N_0$ is the initial number of microbes) is less than $10^{-2}$ for temperatures above 300°C. These facts suggest that microbial survival rate during Mars ejecta formation may be much less than $10^{-2}$ when the target surface is sufficiently rigid. This must be common for the aged surfaces on Mars.

Table 3　Input parameters for basaltic materials used in the 2-D iSALE calculation. Same as Table 2, the detailed descriptions for values listed here can be found in Collins et al. (2016).

| EOS type | ANEOS[1] |
|---|---|
| Material | Basalt[2] |
| Strength model | Rock |
| Poisson ratio | 0.25 |
| Melting temperature (K) | 1360 |
| Thermal softening coefficient | 0.7 |
| Simon parameter A (GPa) | 4.5 |
| Simon parameter C | 3.0 |
| Cohesion (undamaged) (MPa) | 20 |
| Cohesion (damaged) (MPa) | 0.01 |
| Internal friction (undamaged) (MPa) | 1.4 |
| Internal friction (damaged) (MPa) | 0.6 |
| Limiting strength (GPa) | 25 |
| Minimum failure strain | $10^{-4}$ |
| Constant for the damaged model | $10^{-11}$ |
| Threshold pressure for damaged model (MPa) | $3 \times 10^{8}$ |

[1] Thompson and Lauson, 1972; Melosh, 2007.

[2] Modified by Collins and Melosh (2014) after Pierazzo et al. (2005)

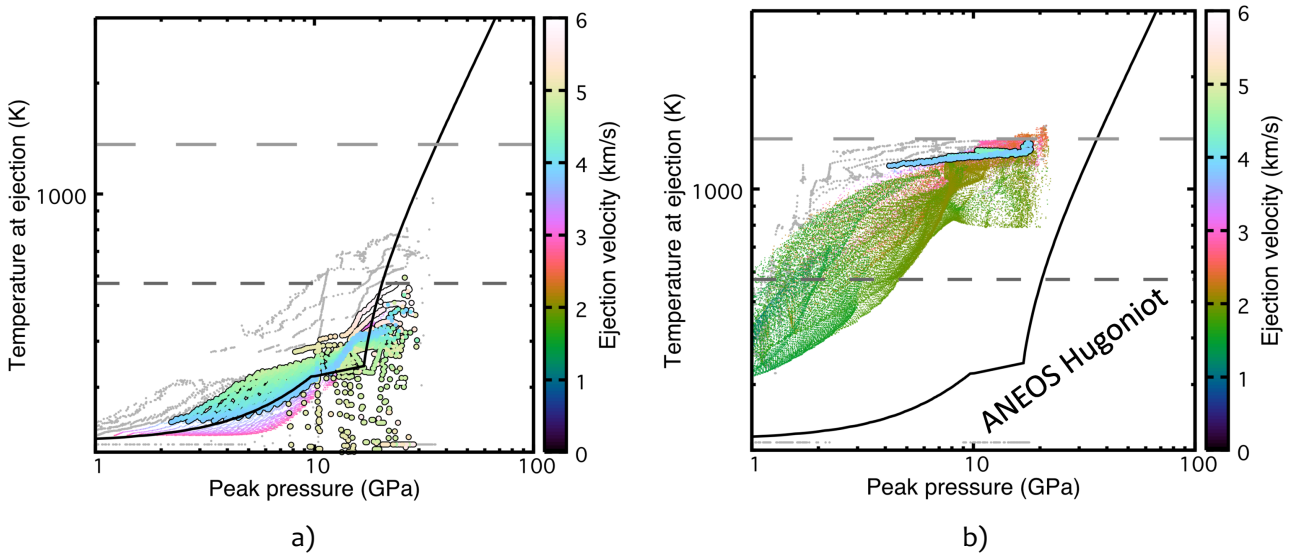

Fig. 5　Temperature of ejected particles at ejection in a hypervelocity impact in relation to ejection velocity; a) without internal friction and plastic deformation, and b) with internal friction and plastic deformation. The two grey dashed lines indicate the temperatures required for the melting of basaltic rocks (upper) and sufficient sterilization (lower).

Park et al. (2008) reported an apparent evidence of sufficient shock heating in Dhofar 378. This fact supports the above argument that the microbial survival rate is sufficiently small in Mars ejecta formation. At the same time, however, Nyquist et al. (2001) reported that a considerable portion of Martian meteorites do not have a sign of shock-heating. This might come from the fact that the geochemical barometers, such as shock melted glasses (e.g., Fritz et al., 2017) and atomic diffusion (e.g., Takenouchi et al., 2017), for the estimation of the temperature rise have relatively low sensitivities at a moderate temperature (several hundred K). Nevertheless, the results by Nyquist et al. (2001) suggest that the Mars ejecta can be sometimes accelerated to the escape velocity and above without sufficient shock-heating. Further studies may be needed on shock-heating and the attainable maximum temperature during Mars ejecta formation. At this moment, the microbial survival rate = 0.1 in Mars ejecta formation is considered to be a good assumption. This assumption seems to be too conservative for the rigid Martian surface material, and reasonable for non-rigid Martian surface material, based on the results of numerical analysis shown in Fig. 5, which may be consistent with the facts of Martian meteorites as well.

## 5. Sterilization by Aerodynamic Heating

Mars ejecta produced by hypervelocity impacts on the Martian surface must undergo aerodynamic deceleration and consequent aerodynamic heating along their flight trajectory in Martian atmosphere. From an orbital mechanics viewpoint, Mars ejecta need to have at least an initial ejection velocity higher than 3.8 km/s to reach the orbits of the Martian moons. The higher the initial ejection velocity, the greater the aerodynamic heating they receive. For Mars ejecta having small diameters, aerodynamic deceleration and aerodynamic heating become remarkable. In order to examine the effect of aerodynamic heating on Mars ejecta ascending to the Martian moons, trajectory calculations under the influence of the Martian atmosphere, CFD computations around Mars ejecta in hypersonic flight, and heat transfer analysis of Mars ejecta are conducted in this study.

### 5.1. Numerical Approaches

For the purpose of this analysis, a Mars atmosphere model, which is shown in Fig. 6, has been developed by using Mars-Global Reference Atmospheric Model 2005 (Mars-GRAM 2005 v1.3) (Duvall et al., 2005; Justus et al., 2005), as a global and annual average of the Martian atmosphere in 2000. At first, a representative aerodynamic drag coefficient and a distribution of aerodynamic heat transfer rate around Mars ejecta are estimated by CFD calculations. In this analysis, Mars ejecta are assumed to be in a spherical form, having diameters ranging from 0.01 to 10 m. Atmospheric temperature and pressure are assumed as 210 K and 0.8 kPa, respectively, which are the values at an altitude of 0 m in the Mars atmosphere model. The Mars atmosphere composition is assumed as 95% $CO_2$ and 5% $N_2$. JAXA Optimized Nonequilibrium Aerothermodynamic Analysis (JONATHAN) code (Fujita et al., 2006), which a compressible flow solver with chemical reactions among high-temperature gases incorporated, is used to compute an aerodynamic drag coefficient and a distribution of aerodynamic heat transfer rate for Mars ejecta.

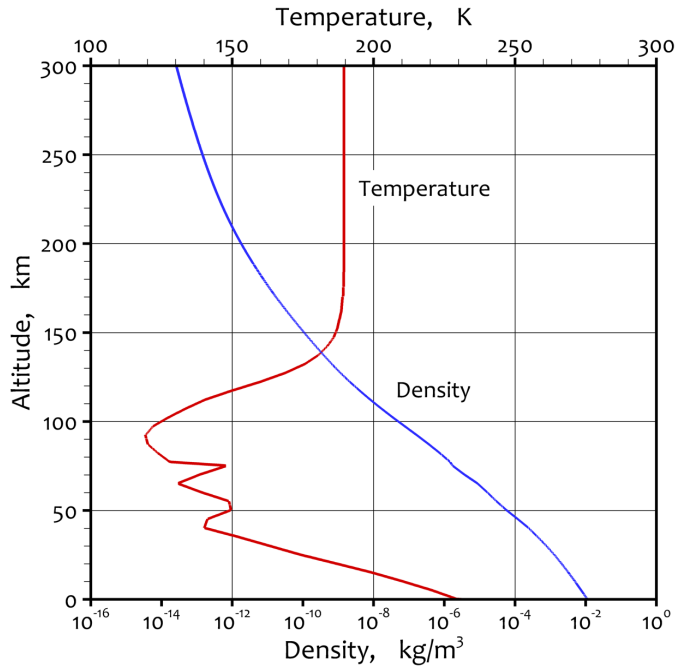

Fig. 6   Mars atmosphere model developed by using Mars-GRAM 2005 v1.3 (Duvall et al., 2005; Justus et al., 2005).

Secondly, based on the aerodynamic drag coefficient obtained above, orbital dynamics of Mars ejecta in the Martian atmosphere is numerically simulated by changing initial ejection angles and velocities of Mars ejecta, to obtain the threshold initial ejection velocity required for Mars ejecta to reach the Phobos orbit (evolution radius = 9,376 km) for each initial ejection angle and diameter of Mars ejecta. In this analysis, a trajectory analysis code used in the preceding study (Fujita et al., 2012) is used. Time history of aerodynamic heat transfer rate at the forebody stagnation point is determined by Tauber's formula (Tauber et al., 1990; Tauber and Sutton, 1991) along the flight trajectory. Distributions of aerodynamic heat transfer rate around Mars ejecta are then estimated by using the representative distribution obtained by the CFD analysis.

According to time history of aerodynamic heat transfer rate distributions around Mars ejecta, unsteady heat conduction analysis about Mars ejecta is conducted to obtain evolution of temperature distribution inside Mars ejecta, with taking radiation cooling on the surface into account. In this analysis, Mars ejecta are assumed as homogeneous basalt having density of 2.8 $g/cm^3$, specific heat of 1.2 J/g·K, thermal conductivity of 2.3 W/m·K, and emissivity of 0.95. The initial temperature of Mars ejecta is assumed as 210 K, which is identical with the atmospheric temperature on the Martian surface. Finally, according to evolution of internal temperature distributions, microbial survival rate is deduced as a function of the initial ejection angle and the diameter of Mars ejecta.

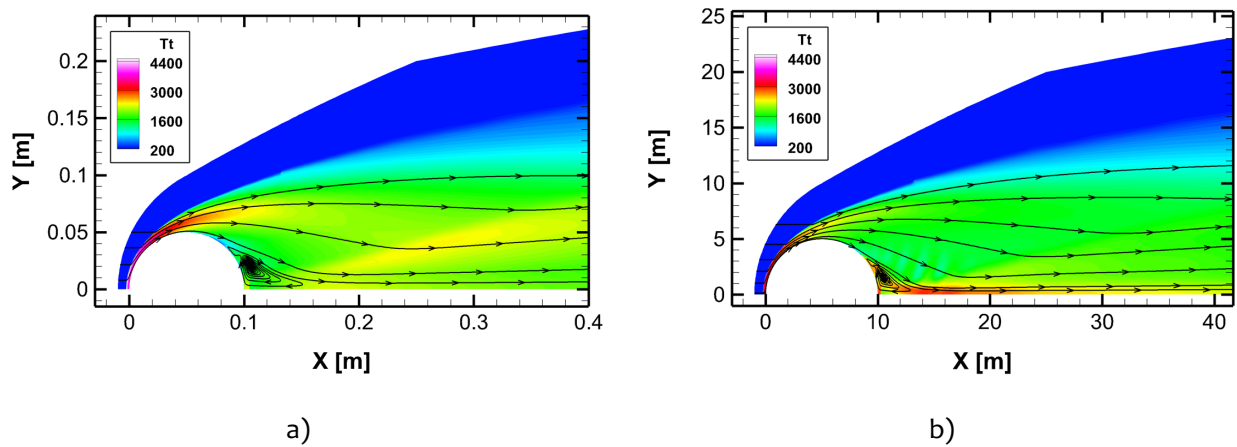

Fig. 7  Distributions of gas temperature and flow streamlines around Mars ejecta obtained by JONATHAN; a) diameter = 0.1 m and b) 10.0 m.

## 5.2. Numerical Results

Distributions of gas temperature and flow streamlines obtained by CFD analysis are presented in Fig. 7 for representative diameters, 0.1 and 10 m. It is seen that gas temperature around Mars ejecta is higher than 1,000 K, suggesting that microbes on the surface are quickly annihilated completely. Gas temperature distribution hardly depends on diameter of Mars ejecta. Distributions of heat transfer rate around Mars ejecta are given in Fig. 8. As a diameter of Mars ejecta increases, the heat transfer rate decreases accordingly. Because of a well-known numerical instability which is inevitable in the axisymmetric analysis, there are some errors in heat transfer rate around the forebody stagnation point. In order to avoid such errors, a fitting function is given to reproduce distribution of the heat transfer rate, as denoted by a solid curve in Fig. 8. The fitting function is used to determine evolution of the heat transfer rate distribution around Mars ejecta from evolution of the heat transfer rate at the stagnation point, which is obtained by trajectory analysis. For conservative analysis of heat conduction, heat transfer to the aftbody region is neglected in the fitting function, as shown in Fig. 8.

The results of trajectory analysis are summarized in Figs.   and 10, in terms of the threshold velocity to reach the Phobos orbit and the corresponding total heat transferred to the stagnation point for Mars ejecta reaching the Phobos orbit, as a function of the ejection angle and the diameter (it should be noted that the ejection angle is denoted as the flight path angle in the figures). The threshold initial velocity becomes high as the ejection angle decreases and as the diameter decreases. This is because Mars ejecta are more decelerated by aerodynamic drag due to increase in flight path length through the Martian atmosphere as the ejection angle decreases. Since the aerodynamic drag force is proportional to the square of the diameter, Mars ejecta having small diameters are more likely to be decelerated than those having large diameters. As a result, Mars ejecta having diameters smaller than 0.1 m are difficult to leave the Martian atmosphere. The total heat transferred to the stagnation point of Mars ejecta becomes significantly high as the ejection angle decreases and as the diameter decreases. This is because the initial velocity has to be increased as described above, and because the stagnation-point heat transfer rate essentially increases as the radius of local curvature decreases (e.g. Tauber and Sutton, 1991). More detailed analysis on thermal behavior of Mars ejecta is given in the next section.

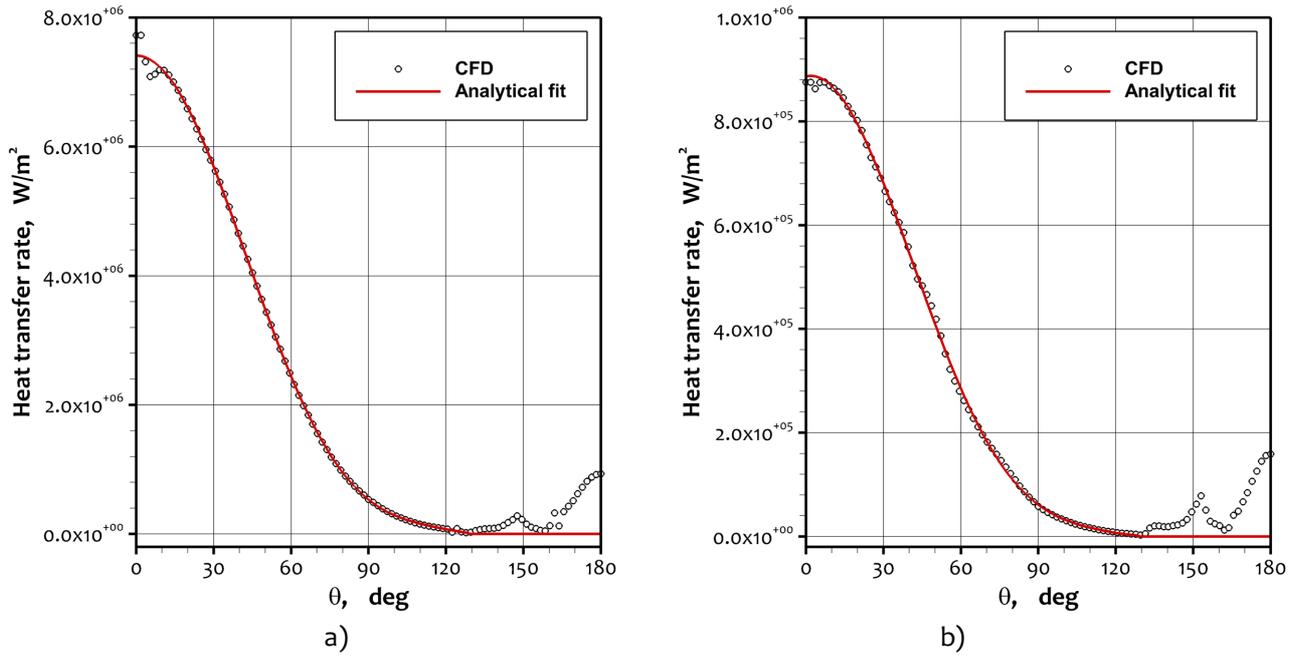

Fig. 8 Distributions of heat transfer rate around Mars ejecta obtained by JONATHAN; a) radius = 0.1 m and b) 10.0 m.

### 5.3. Internal Temperature of Mars Ejecta

According the results obtained by CFD and trajectory calculations, time history of the aerodynamic heat transfer distribution around Mars ejecta is obtained. Using this as the boundary condition, an unsteady differential equation of heat conduction in a 2-D axisymmetric form is integrated with respect to time by the 4$^{th}$ order Runge-Kutta method. Evolution of internal temperature distribution so obtained is illustrated in Fig.   as a representative example. In this case, since aerodynamic heating is completed within several ten seconds from ejection, the region with the highest temperature is located near the surface in Fig. a, then gradually diffuses to the whole sphere over time. The maximum temperature encountered at each internal location during the period from ejection to arrival at the Phobos orbit is shown in Fig.   for the representative diameters of Mars ejecta. It is seen that, as the diameter decreases, a portion of the region that encounters temperature above 500°C quickly increases, and the maximum temperature becomes higher.

### 5.4. Microbial Survival Rate for Aerodynamic Heating

Finally, microbial survival rates for aerodynamic heating are estimated from the maximum internal temperature obtained above, with an assumption that microbes are completely annihilated if temperature is maintained above 500°C for more than 0.5 sec, which is a widely accepted criterion for microbial sterilization. The results are summarized in Table 4. It is seen that the most portion of microbes are expected to be annihilated for Mars ejecta with diameters smaller than 0.03-m, while they are scarcely sterilized for diameters as large as 0.1 m and above. As already described above, Mars ejecta having diameters less than 0.1 m are difficult to reach the Phobos orbit due to aerodynamic deceleration (see Sec. 5.2). Above all, it is concluded that most of Mars ejecta transported to the Martian moons have diameters as large as 0.1 m and more, and are hardly sterilized except for their surfaces. Those whose diameter is smaller than 0.1 m hardly reach the Martian moons, and are considerably sterilized even if they arrive.

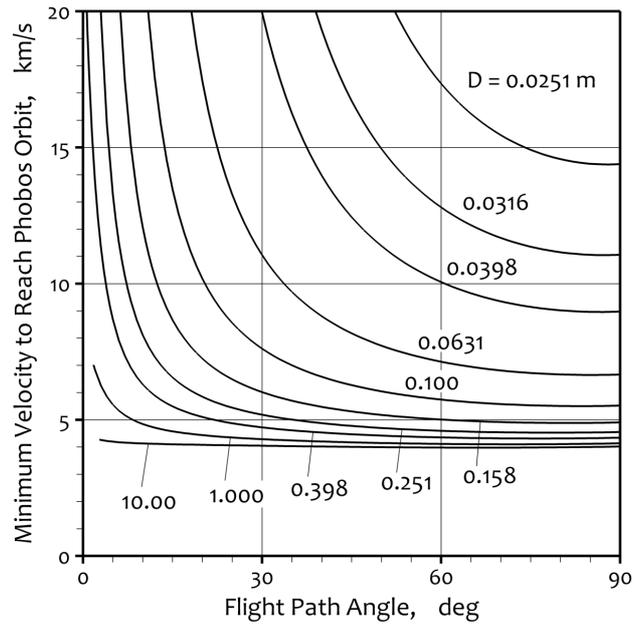

Fig. 9 Threshold initial velocity of Mars ejecta to reach Phobos orbit for different ejection angles and diameters.

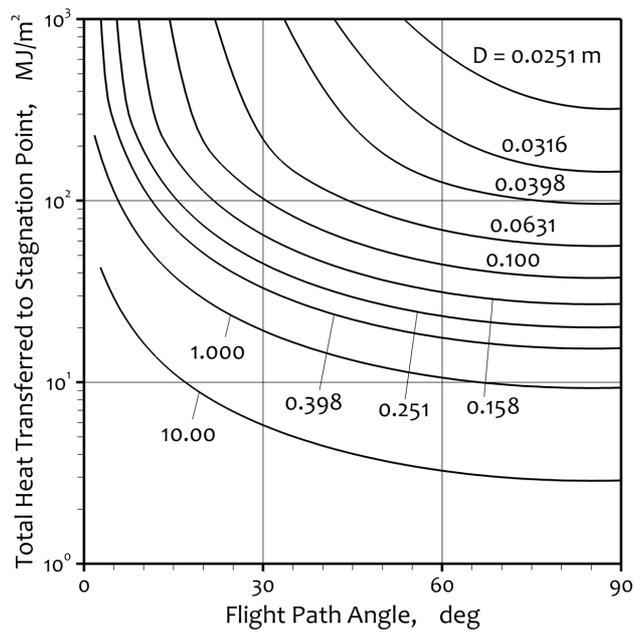

Fig. 10 Total heat transferred to stagnation point of Mars ejecta for different ejection angles and diameters.

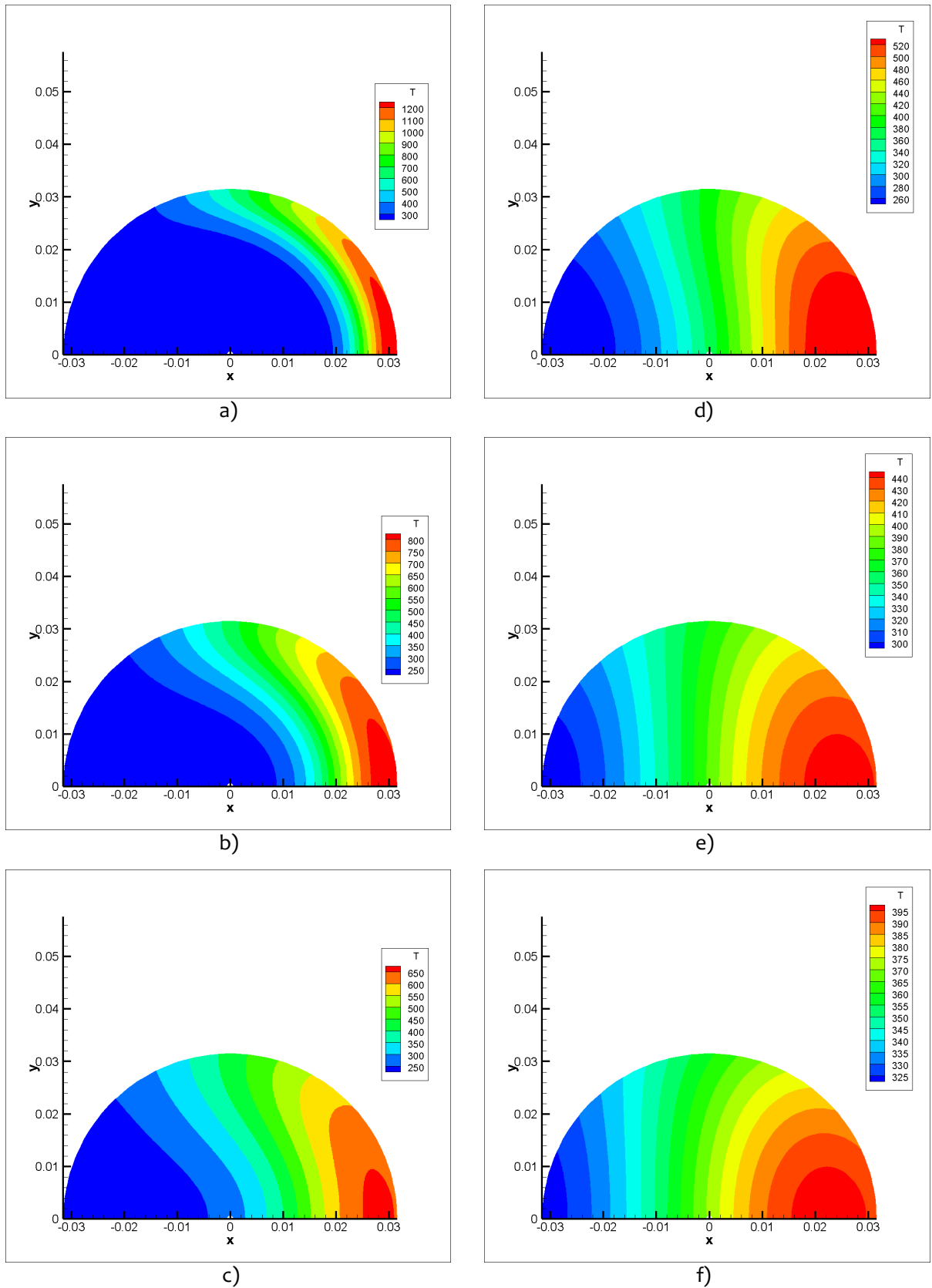

Fig. 11 Evolution of internal temperature of Mars ejecta with diameter = 0.0631 m and ejection angle = 90°; x- and y-axis show spatial distance from the center of ejecta in m, and the positive direction of x-axis is the direction of flight; a) *t* = 20 sec., b) 50 sec., c) 100 sec., d) 200 sec., e) 400 sec., and f) 600 sec.

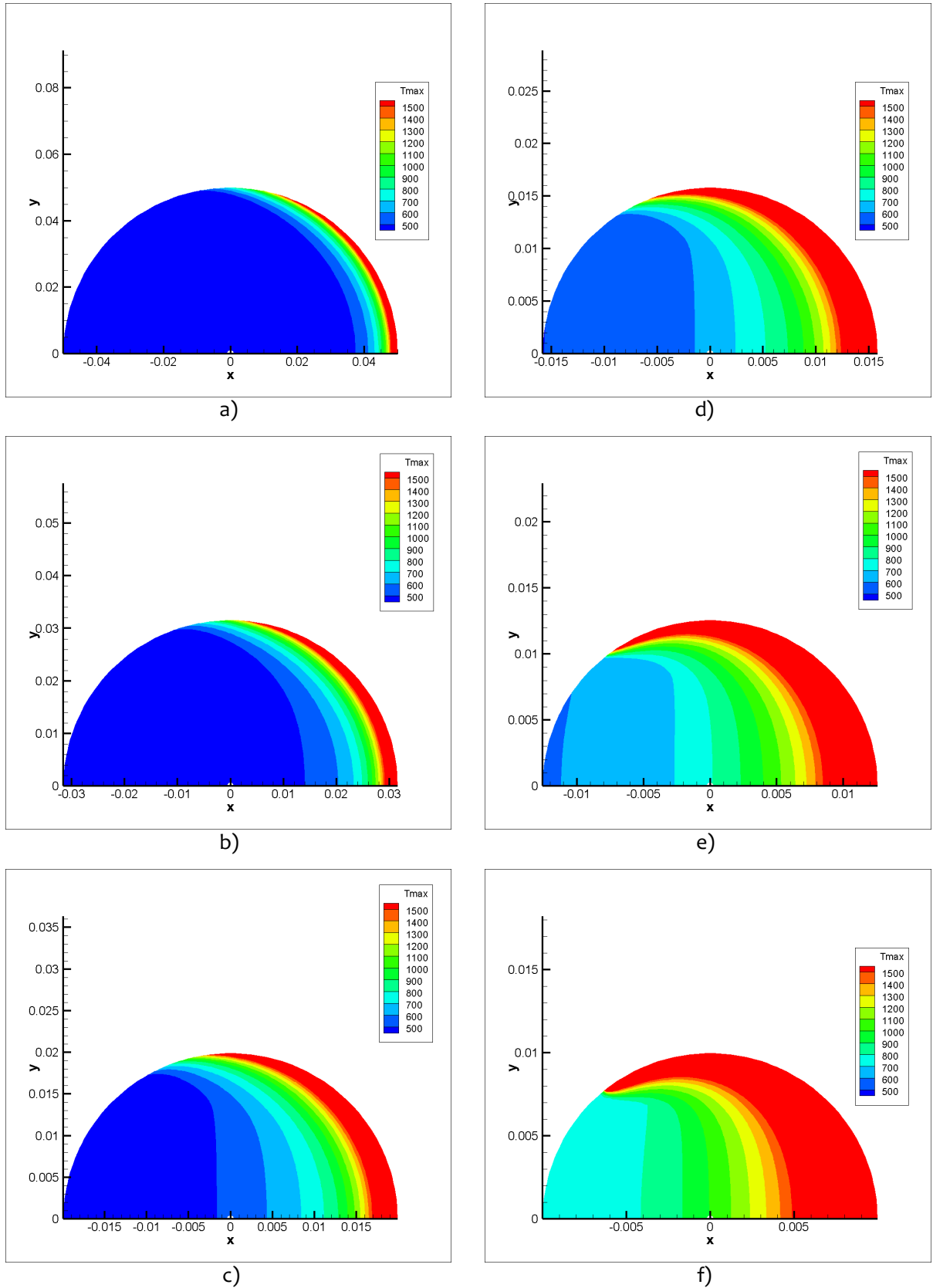

Fig. 12  Maximum internal temperature of Mars ejecta with ejection angle = 90°; x- and y-axis show spatial distance from the center of ejecta in m, and the positive direction of x-axis is the direction of flight; a) diameter = 0.1 m, b) 0.0631 m, c) 0.0398 m, d) 0.0316 m, e) 0.0251 m, and f) 0.0200 m.

Table 4  Microbial survival rate for aerodynamic heating.

| Diameter (m) | Flight path angle (deg) | | | |
|:---:|:---:|:---:|:---:|:---:|
| | 90° | 75° | 60° | 45° |
| 0.0100 | 0.000 | 0.000 | 0.000 | 0.000 |
| 0.0158 | 0.000 | 0.000 | 0.000 | 0.000 |
| 0.0200 | 0.138 | 0.123 | 0.067 | 0.000 |
| 0.0251 | 0.323 | 0.310 | 0.266 | 0.190 |
| 0.0316 | 0.494 | 0.481 | 0.431 | 0.341 |
| 0.0398 | 0.632 | 0.625 | 0.590 | 0.507 |
| 0.0631 | 0.806 | 0.803 | 0.785 | 0.748 |
| 0.100 | 0.899 | 0.897 | 0.888 | 0.867 |
| 0.398 | 0.987 | 0.987 | 0.986 | 0.983 |
| 1.000 | 0.997 | 0.997 | 0.997 | 0.996 |

## 5.5. Effect of Blast Flows

The aerodynamic heating described above is accurate as long as Mars ejecta undergo a single hypersonic flight in the stationary Martian atmosphere. However, in the presence of blast flows generated by a gigantic meteoroid impact on the Martian surface, even though the Mars ejecta having small diameters do not have initial velocities higher than 3.8 km/s to reach the Phobos orbit, they may be easily accelerated by blast flows and travel with them toward the orbit of Martian moons. In such a case, aerodynamic heating may be reduced since the shock layer is not generated around Mars ejecta or the generated shock layer will be weaker than that generated in a single hypersonic flight.

To assess the heat transfer rate in such a situation, a one-dimensional shock wave problem was solved for the Martian atmosphere to assess the thermodynamic properties of blast flows behind the preceding shock wave. It is found that, in order to generate blast flows faster than 3.8 km/s in the Martian atmosphere, it is necessary for the blast flow temperature and the pressure to be higher than 5,000 K and 10 kPa, respectively, even though dissociation of $CO_2$ and $N_2$ are taken into consideration. The cold-wall heat flux transferred to Mars ejecta in such a hot gas is expected to be at the order of 200 kW/m$^2$ or higher. Flight time of Mars ejecta across the Martian atmosphere is in general 20 seconds or longer for the flight velocity of 3.8 km/s. The maximum temperature of Mars ejecta after 20-sec heating at 200 kW/m$^2$ is estimated as shown in Table 5. In this bulk thermal balance calculation, density and heat capacity of Mars ejecta are assumed to be 2.8 g/cm$^3$ and 1.2 J/g·K, respectively, and the initial temperature is assumed to be 210 K. The results suggest that small Mars ejecta are more likely to be sterilized, while those having diameters greater than 0.03 m are scarcely sterilized.

Based on the results of both Tables 4 and 5, it is concluded that Mars ejecta of 0.01 m in diameter and smaller do not reach the Phobos orbit without complete sterilization due to aerodynamic heating. Those having diameters between 0.01 and 0.03 m are expected to be partly sterilized. However, it should be noted that Mars ejecta of 0.1 m in diameter and smaller are essentially not expected to reach the Phobos orbit because of aerodynamic deceleration in the Martian atmosphere, as shown in Fig. . Even though they are initially accelerated by the blast flow, they will be decelerated as well when the blast flow slows down during expansion from the center of the meteoroid impact point. For example, a Mars ejecta of 0.01 m in diameter suffers from deceleration of 30 m/s² even if the difference between its flight velocity and the blast flow velocity reaches only 100 m/s at the blast flow pressure of 10 kPa. This means that such small Mars ejecta quickly follow the motion of the surrounding atmosphere. They will slow down as the blast flow velocity decreases during expansion and hardly escape from the Martian atmosphere.

Table 5   Estimated maximum temperature of Mars ejecta after 20-sec heating at 200 kW/m² (in blast flows).

| Diameter (m) | Bulk heating (kJ) | ΔT (K) | T (K) | Expected sterilization |
|---|---|---|---|---|
| 0.0100 | 1.257 | 800.0 | 1,010 | Completed |
| 0.0158 | 3.137 | 506.3 | 716.3 | Mostly done |
| 0.0200 | 5.027 | 400.0 | 610.0 | Fairly done |
| 0.0251 | 7.917 | 318.7 | 528.7 | Partly done |
| 0.0316 | 12.55 | 253.2 | 463.2 | Scarcely done |

## 6.  Conclusions

Potential microbial density on the Martian surface was deduced from a thorough survey of the literatures on microbial density measured in the terrestrial areas having the similar arid and cold environments as the Martian surface. As an asymptotic model of the arid and the cold limit, microbial density on the Martian surface is likely to range from $10^6$ to $10^9$ cells/kg. Based on this estimation, a probabilistic model is defined by $P_{d,m} = \sqrt{2/\pi}\exp[-2(x-7.5)^2]$ with $x = \log_{10} n$ where $n$ is the microbial density in cells/kg. Microbial survival rate during Mars ejecta formation by a meteoroid impact on the Martian surface is estimated to be 0.1, based on the results of impact analysis with and without taking account of internal friction and plastic deformation of the colliding meteoroid and the target ground. Trajectory calculations of departing Mars ejecta show that the ejecta smaller than 0.1 m in diameter are difficult to escape from the Martian atmosphere. Thermal analysis along the flight trajectory shows that such small ejecta, even though they reach the Phobos orbit because of high initial velocities, are mostly sterilized due to aerodynamic heating. The Mars ejecta smaller than 0.02 m in diameter are completely sterilized when the ejection angle is 45°. Acceleration of the Mars ejecta by the blast flow was found to potentially reduce the microbial sterilization rate. However, because of high temperature and pressure of the blast flow, the Mars ejecta having small diameters are mostly sterilized by aerodynamic heating as well. Above all, the general characteristics of Mars ejecta departing from the Martian atmosphere and the baseline dataset of microbial density in Mars ejecta departing for Martian moons have been obtained for Paper 2 where statistical assessment of microbial contamination probability is conducted. Our results imply that a considerable fraction of microbes living on Mars continue to drain to the space due to periodic meteoroid impacts as long as our conservative assumptions about the number density of the potential microbes and the survival rate during the launch are valid. The fate of the transported microbes on the Martian moons are addressed in detail in Paper 2.

## Acknowledgements

A part of this research was conducted by using the iSALE computer program. We wish to acknowledge the developers of iSALE, including G. Collins, K. Wünnemann, B. Ivanov, J. Melosh, and D. Elbeshausen.